\documentclass[twocolumn]{article}

\usepackage{cite}            
\usepackage{newtxtext}        
\usepackage[euler]{textgreek} 
\usepackage{lipsum}	          
\usepackage[svgnames]{xcolor} 
\usepackage[utf8]{inputenc}   
\usepackage[T1]{fontenc}      
\usepackage[shortlabels]{enumitem}  	
\usepackage{marginnote}       

\usepackage{amsmath}    	
\usepackage{amssymb}	    
\usepackage{amsfonts}		  
\usepackage{amsthm}    	  
\usepackage{mathtools}		
\usepackage{newtxmath} 	  
\usepackage{bm}           
\usepackage{cases}     	

\usepackage[caption=false,font=footnotesize]{subfig}
\usepackage{graphicx}	          
\usepackage{booktabs}   	      
\usepackage{siunitx}            
\usepackage[export]{adjustbox}  
\usepackage{mdframed}		        
\usepackage{floatrow}		
\usepackage{subfig}	    

\usepackage{geometry}		    
\usepackage{fancyhdr}       
\usepackage{titling}        
\usepackage{titlesec}        
\usepackage[keeplastbox, nospread]{flushend}	
\usepackage{abstract}		    
\usepackage[hidelinks]{hyperref}		    
\usepackage{datetime}		    

\numberwithin{equation}{section}	
\graphicspath{{figures/}}         
\setlist{noitemsep,topsep=0.5ex} 		          
\definecolor{myTeal}{rgb}{0.0, 0.45, 0.55}
\definecolor{myRed}{rgb}{0.45, 0.0, 0.0}
\newcommand{\accentcolor}{myTeal}
\newcommand{\lightaccentcolor}{gray!15}
\newdateformat{mydateformat}{\THEDAY \ \monthname[\THEMONTH] \THEYEAR}

\allowdisplaybreaks          

\titleformat{\part}[display]{\centering\sffamily\Huge\bfseries}{\textsc{\partname}\thepart\\ \vspace*{1ex}{\Huge\color{\accentcolor}------ $\cdot$ ------}}{1.2ex}{\huge}
\titleformat{\chapter}{\sffamily\bfseries\Huge\color{\accentcolor}}{\thechapter \	}{10pt}{}
\titleformat*{\section}{\Large\sffamily\color{\accentcolor}}
\titleformat*{\subsection}{\large\sffamily\color{\accentcolor}}
\titleformat*{\subsubsection}{\sffamily\itshape\color{\accentcolor}}

\let\oldheadrule\headrule
\let\oldfootrule\footrule
\renewcommand{\headrule}{{\color{\accentcolor}\oldheadrule}}%
\renewcommand{\footrule}{{\color{\accentcolor}\oldfootrule}}%
\fancypagestyle{firstpage}{ 
	\fancyhf{}
	\lhead{\sffamily \footnotesize }
	\chead{}
	\rhead{\sffamily \footnotesize \mydateformat\today}
	\lfoot{\sffamily \footnotesize }
	\cfoot{}
  \rfoot{}
}

\newcommand{\T}{^{\mathsf{T}}}

\newcommand{\B}[1]{\if#1\relax\bm{#1}\else\mathbf{#1}\fi} 
\newcommand{\R}[1]{\mathrm{#1}}						      
\newcommand{\C}[1]{\mathcal{#1}}
\newcommand{\BB}[1]{\mathbb{#1}}

\newtheoremstyle{mythmstyle}  
{1em}
{1em}
{\itshape}
{}
{\sffamily \color{\accentcolor}}
{.}
{ }
{{\bfseries\thmname{#1}\thmnumber{ #2}}\thmnote{ (#3)}}  

\theoremstyle{mythmstyle}			

\mdfdefinestyle{myshadedthm}{backgroundcolor=\lightaccentcolor,linewidth=0pt,innerleftmargin=2ex,innerrightmargin=2ex,innertopmargin=0ex}
\newmdtheoremenv[style=myshadedthm]{theorem}{Theorem}[section]
\newmdtheoremenv[style=myshadedthm]{definition}[theorem]{Definition}
\newmdtheoremenv[style=myshadedthm]{assumption}[theorem]{Assumption}
\newmdtheoremenv[style=myshadedthm]{proposition}[theorem]{Proposition}
\newmdtheoremenv[style=myshadedthm]{lemma}[theorem]{Lemma}
\newmdtheoremenv[style=myshadedthm]{corollary}[theorem]{Corollary}
\newmdtheoremenv[style=myshadedthm]{remark}[theorem]{Remark}
\newmdtheoremenv[style=myshadedthm]{conjecture}[theorem]{Conjecture}
\newtheorem{problem}[theorem]{Problem}

\title{\color{\accentcolor}{\huge \sffamily Minimax Flow over Acyclic Networks:\\Distributed Algorithms and Microgrid Application}}
\author{\color{\accentcolor}{\normalsize \sffamily \itshape 
Marco Coraggio, 
Saber Jafarpour,
Francesco Bullo*,
Mario di Bernardo*}
\thanks{This work was in part supported by the Research Project PRIN 2017 “Advanced Network Control of Future Smart Grids” funded by the Italian Ministry of University and Research (2020–2023)--http://vectors.dieti.unina.it, {\color{black}and by the} AFOSR grant FA9550-22-1-0059.}
\thanks{%
M.~Coraggio is with the Scuola Superiore Meridionale (SSM), School for Advanced Studies (marco.coraggio@unina.it).
S.~Jafarpour is with the Dept. of Electrical and Computer Engineering, Georgia {\color{black}Inst.} of Technology (saber@gatech.edu).
F.~Bullo is with the Dept. of Mechanical Engineering, {\color{black}Univ.} of California Santa Barbara.
M.~di Bernardo is with the Dept. of Information Technology and Electrical Engineering, {\color{black}Univ.} of Naples Federico II, and with the {\color{black}SSM} (mario.dibernardo@unina.it).
}
\thanks{*These authors contributed equally.}}
\date{}
\renewcommand\footnotemark{}

\makeatletter
\ifx\@date\empty
\patchcmd{\@maketitle}{{\@bspredate \@date \@bspostdate} \maketitlehookd \par \vskip 1.5em}{\vskip 0.5em}{}{}
\fi
\makeatother

\begin{document}

\maketitle
\thispagestyle{firstpage} 
\begin{abstract}
\noindent \normalsize {\color{\accentcolor}{\textsf{\textbf{Abstract. \;}}}}%
\bf{Given a flow network with variable suppliers and fixed consumers, the minimax flow problem consists in minimizing the maximum flow between nodes, subject to flow conservation and capacity constraints.
We solve this problem over acyclic graphs in a distributed manner by showing that it can be recast as a consensus problem between the maximum downstream flows, which we define here for the first time.
Additionally, we present a distributed algorithm to estimate these quantities.
Finally, exploiting our theoretical results, we design an online distributed controller to prevent overcurrent in microgrids consisting of loads and droop-controlled inverters.
Our results are validated numerically on the CIGRE benchmark microgrid.}
\end{abstract}


\section{Introduction}\label{sec:introduction}

\subsubsection*{Problem description and motivation}
Flow networks are dynamical systems where a commodity of interest is provided by supplier nodes, flows over the network edges, and reaches consumer nodes.
Critical infrastructure networks such as power grids, water distribution networks, and traffic networks are modeled as flow networks, with the commodity of interest being electrical power, water, and vehicles, respectively~\cite{ARB-DJH:81,JB-BG-MCS:09,EL-GC-KS:14}. 
A fundamental problem in these networks is {\color{black}to cater for} consumers' demands, while keeping the commodity flows over the network edges below their maximum capacities. 
Hence, a valuable optimization problem is to minimize the maximum flow over the network edges, thereby ensuring that no edge capacity is exceeded.
Violation of capacity constraints is a safety-critical event, with a potential to cause disruptions or faults in real-world infrastructure networks.
Typically, the resulting minimax flow problem is solved offline in a centralized fashion, so that the ``right'' flows can be assigned to {\color{black}the} network edges.
{\color{black}However}, recent changes in infrastructure networks, due to the increase in demand, the integration of numerous smart devices and the need for higher energy efficiency, have shown the limitations of such centralized approaches.

In this paper, we propose a distributed solution to the minimax flow problem over acyclic networks {\color{black}consisting of suppliers} and consumer nodes, where the former can adjust their supply rates to satisfy fixed consumption demands in the latter. 
In particular, by solving a distributed consensus problem, we propose a strategy for supplier generation that minimizes the maximum flow over all edges, subject to flow conservation and safety constraints. 
As a case study of relevance in applications, we apply our distributed approach to AC microgrids consisting of resistive loads and droop-controlled distributed energy units. 
We show that {\color{black}our algorithm} is an effective solution to adjust the suppliers' generation rates in order to prevent overcurrents on the network edges while fulfilling the demands of the consumers.

\subsubsection*{Literature on network optimization problems}%
One of the earliest formulations of minimax optimization problem{\color{black}s on graphs} is the \emph{minimax location problem} \cite{hakimi1964optimum}, where the objective function is the distance between a facility node to be placed in the network and the other nodes in the graph.
Later studies on this topic include~\cite{okelly1991solution,campbell2006upgrading}.
In~\cite{hammer1969timeminimizing,garfinkel1971bottleneck}, the \emph{time-minimizing transportation problem} was studied, where source nodes and sink nodes are two disjoint sets making up a bipartite graph, and the objective is to minimize the maximum transportation time among all utilized edges. 
In~\cite{ichimori1980finding}, the \emph{minimax transportation problem} is introduced for cyclic graphs with one source node and one sink node, with the objective of minimizing the maximum flow in the network. 
Later, in~\cite{ahuja1986algorithms}, the problem is recast as a linear program and several solution algorithm are presented.

Surprisingly, to the best of our knowledge, relatively few distributed solutions of {\color{black}minimax problems on graphs} have been presented in the existing literature (see \cite{nedic2018distributed} for a recent review of {\color{black}distributed network} optimization algorithms).
Examples of existing distributed approaches, {\color{black} although not applicable to minimax \emph{flow} problems,} include those presented in {\color{black} \cite{gharesifard2013distributed}, where two networks are in competition to maximize and minimize an objective function, and \cite{yang2019cooperative}, where agents are divided into two groups for computing two continuous decision variables in a minimax optimization.}
For the specific case of flow networks, a Newton-based distributed algorithm is presented in  \cite{jadbabaie2009distributed} for minimizing the sum of all flows, while an accelerated algorithm for a similar problem is described in \cite{zargham2014accelerated}.
Also, a distributed algorithm for minimizing the $p$-norm of flows was presented in~\cite{anaraki2011acceleration}, which approximates the minimax flow problem when $p$ becomes very large.

\subsubsection*{Literature on microgrid protection}
Protection against faults (such as overcurrents) in microgrids can be ensured through three kinds of interventions: \emph{prevention} (before the unwanted events), \emph{detection} (during the events), and \emph{management} (right after the events).
{\color{black}In the literature, most studies} focus on detection and management (see \cite{memon2015critical,hooshyar2017microgrid,hosseini2016overview} and references therein). 
However, fault prevention is one area in which the use of intelligent control strategies could prove particularly fruitful, given the many challenges with fault detection and management algorithms currently available for microgrids \cite{khederzadeh2018identification,nahata2017decentralized,goyal2014overload}.

An optimization problem to find the maximum permissible loading is solved in \cite{khederzadeh2018identification} through genetic algorithms{\color{black}, to} prevent the occurrence of cascading failures. Overvoltages are prevented in \cite{nahata2017decentralized} via a decentralized control scheme that curtails the active power output of the generators when necessary, while a control strategy is presented in \cite{goyal2014overload} to prevent overloading of distributed generators during peak demand time, employing battery storage units that can intervene smoothly. 
Further distributed control strategies for microgrids include \cite{etemadi2012decentralized,shah2011decentralized,nianniancai2010decentralized,anand2013reducedorder,gu2014modeadaptive,simpson-porco2013synchronization} but are not specifically aimed at solving minimax {\color{black}problems}. 
A minimax optimization problem for networks of microgrids is solved in a distributed fashion in \cite{bersani2017distributed}, {\color{black}minimizing} a function of the energy stored in the microgrids and the power flows between them, {\color{black}controlling} the latter. 

\subsubsection*{Contributions}%
The key contributions of this paper can be summarized as follows:
\begin{enumerate}
    \item we establish a connection between solving the minimax {\color{black}flow} problem over an acyclic graph and achieving consensus of the \emph{maximum downstream flows}, that we define here for the first time;
    \item we propose a distributed estimation strategy to evaluate the maximum downstream flows of a network of interest;
    \item we exploit our theoretical results and {\color{black}an} estimation strategy to obtain an online distributed controller to minimize the maximum power flow on the lines of a microgrid, by adjusting dynamically the power generated by the suppliers, thus preventing overcurrents in the grid.
\end{enumerate}

When compared to the existing literature, our objectives and methodology are closer in flavor to those presented in~\cite{anaraki2011acceleration}, with the important differences that therein (i) consumers {\color{black}can} absorb any amount of commodity and (ii) only an approximate solution of the minimax flow problem is obtained.
All the other references we reviewed differ from our work in major aspects, such as the optimization problem (e.g., minisum rather than minimax, as in \cite{zargham2014accelerated}) or the network structure (e.g., single source and single sink, with cyclic graphs, as in \cite{ichimori1981weighted}).





\section{Review of flow networks}
\label{sec:review}
\subsubsection*{Notation}%

We let {\color{black}$\max(\varnothing) = 0$.}
Letting $\C{Q}$ and $\C{R}$ be sets, 
$\lvert \C{Q} \rvert$ is the cardinality of $\C{Q}$,
and $\C{Q} \rightrightarrows \C{R}$ {\color{black}is} an application from $\C{Q}$ to all subsets of $\C{R}$. 
Given a matrix $\B{A}$, $\R{ker}(\B{A})$ is its null space (kernel), and {\color{black}$\B{A}^\dagger$} is its Moore-Penrose (pseudo-)inverse {\color{black}\cite{penrose1955generalized}}.

\subsubsection*{Graph theory}%
Letting $\C{G} = (\C{V}, \C{E})$ be a graph, $\C{V}$ and $\C{E}$ are the set of vertices and the set of edges, respectively; $N \triangleq \left| \C{V} \right|$ and $N_\C{E} \triangleq \left| \C{E} \right|$ being the numbers of vertices and edges.
We denote an undirected edge connecting vertices $i$ and $j$ as $\{i, j\}$, and a directed edge from $i$ to $j$ as $(i, j)$.
$\B{A}$ and $\B{L}$ are the \emph{adjacency} and \emph{Laplacian matrices} associated to $\C{G}$.
In an undirected graph, we let $\C{Q}$ be the set of edges in $\C{E}$, after they have been enumerated and oriented in an arbitrary way, and let $\B{B}$ be the \emph{incidence matrix} associated to the graph $(\C{V}, \C{Q})$.
{\color{black}In} a (directed) graph, a (\emph{directed}) \emph{path} is an ordered sequence of vertices such that any pair of consecutive vertices is an edge in the graph.
In a directed graph $(\C{V}, \vec{\C{E}})$, the \emph{out-tree} of vertex $i \in \C{V}$ is the union of all directed paths starting from $i$; moreover, the \emph{out-neighborhood} of a vertex $i$ is the set of all vertices $j$ such that a directed edge $(i, j)$ exists in $\vec{\C{E}}$.

\subsubsection*{Flow networks}
Consider a \emph{flow network} associated to an \emph{undirected} \emph{acyclic} \emph{unweighted} graph $\C{G} = (\C{V}, \C{E})$. 
We define $\C{V}_\R{s} \subset \C{V}$ as the set of \emph{supplier} vertices and $\C{V}_\R{c} \subset \C{V}$ as the set of \emph{consumer} vertices, with $\{\C{V}_\R{s}, \C{V}_\R{c}\}$ being a partition of $\C{V}$. 
Additionally, we let $N_\R{s} \triangleq \left| \C{V}_\R{s} \right| \ge 2$ and $N_\R{c} \triangleq \left| \C{V}_\R{c} \right| \ge 1$ be the number of supplier and consumer vertices, respectively.

\subsubsection*{Commodity}%
We let $m_i \in \BB{R}$ be the amount of \emph{commodity} supplied {\color{black}($m_i > 0$)} or consumed {\color{black} ($m_i \le 0$)} at vertex $i$, and define $\B{m} \triangleq [m_i]_{i \in \C{V}} \in \BB{R}^N$ and $\B{m}_\R{s} \triangleq [m_i]_{i \in \C{V}_\R{s}} \in \BB{R}^{N_\R{s}}$.
We assume that the amounts of consumed commodity ($m_i, i \in \C{V}_\R{c}$) are given, whereas the amounts of supplied commodity ($m_i, i \in \C{V}_\R{s}$) can be controlled, provided that $\B{m}_{\R{min}} \le \B{m}_\R{s} \le \B{m}_{\R{max}}$, where $\B{m}_{\R{min}}, \B{m}_{\R{max}} \in \BB{R}_{>0}^{N_\R{s}}$ are vectors of positive real numbers.%
\footnote{\label{fn:uncontrollable_suppliers}\color{black}If a supplier $i$ is not controllable, it is possible to set $m_{\R{min}, i} =m_{\R{max},i}$.}


\subsubsection*{Flows}
For all $\{i, j\} \in \C{E}$, we let $f_{ij} \in \BB{R}$ denote the \emph{flow} of commodity from $i$ to $j$; $f_{ij} > 0$ if commodity flows from $i$ to $j$ and vice-versa, and $f_{ji} = - f_{ij}$.  
We also define $\B{f} = [f_{ij}]_{(i,j) \in \C{Q}} \in \BB{R}^{N_\C{E}}$.
The flows satisfy the balancing equations
\begin{equation}{\label{eq:flow_network}}
	\sum_{j : \{i, j\} \in \C{E}}f_{ij} = m_i, \quad \forall i \in \C{V},
\end{equation}
which can be written in a more compact form as
\begin{equation}\label{eq:flow_network_compact}
  \B{B} \B{f} = \B{m}.
\end{equation}
Finally, we let $\bar{f}_{ij} \in \BB{R}_{>0}$ be the \emph{capacity} (i.e., maximum flow allowed) of edge $\{i, j\}$, and define $\bar{\B{f}} = [\bar{f}_{ij}]_{(i,j) \in \C{Q}} \in \BB{R}_{>0}^{N_{\C{E}}}$.

Next, we present a result characterizing flows over acyclic networks.
For completeness' sake, we include a short proof.

\begin{lemma}[Flows~\cite{DFD-MC-FB:13}]\label{lem:flows}
	In an acyclic unweighted undirected flow network with incidence {\color{black}matrix} $\B{B}$, Laplacian {\color{black}matrix} $\B{L}$,  and commodity vector $\B{m}$, the flows $\B{f}$ are uniquely determined by commodity conservation {\color{black}\eqref{eq:flow_network_compact}} and are given by
	\begin{equation}\label{eq:flows}
		\B{f} = \B{B}\T \B{L}^\dagger \B{m}. 
	  \end{equation}		  
\end{lemma}

\begin{proof}
	{\color{black}From \cite{simpson-porco2013synchronization}, we have $\B{L}^\dagger \B{L} = \B{I} - \frac{1}{N}\B{1}\B{1}\T$, and, as the graph is unweighted, $\B{L} = \B{B} \B{B}\T$ \cite[Chapter~9]{bullo2020lectures}.
	Then, consider the following expression: $\B{B}\T \B{L}^\dagger \B{B} \B{B}\T = \B{B}\T \B{L}^\dagger \B{L} = \B{B}\T (\B{I} - \frac{1}{N}\B{1}\B{1}\T ) = \B{B}\T$.}
	As the graph is acyclic, {\color{black}$\R{ker} (\B{B}) = \varnothing$} \cite{bullo2020lectures}, and thus
	$
		\B{B}\T \B{L}^\dagger \B{B} = \B{I}.
    $
	Therefore, premultiplying \eqref{eq:flow_network_compact} by $\B{B}\T \B{L}^\dagger$, we get the thesis.
\end{proof}

\section{Problem formulation}%
\label{sec:problem_formulation}


\subsection{Minimax flow problem}%
\label{sec:minimax_problem}


We start by defining the flow safety margin of a network. 

\begin{definition}[Flow safety margin]\label{def:flow_safety_margin}
Given a flow network over $\C{G} = (\C{V}, \C{E})$ with supplied commodity $\B{m}_{\R{s}}$, flows $f_{ij}$ and capacities $\bar{f}_{ij}$, the \emph{flow safety margin} $J_{\C{E}_{\R{r}}} : \BB{R}^{N_\R{s}} \rightarrow \BB{R}_{\ge 0}$, with respect to a given edge set $\C{E}_{\R{r}} \subseteq \C{E}$ is
\begin{equation}\label{eq:J}
    J_{\C{E}_{\R{r}}}(\B{m}_\R{s}) \triangleq \max_{\{i,j\} \in \C{E}_{\R{r}}} \frac{ \left| f_{ij} \right|  }{\bar{f}_{ij}}.
\end{equation}
\end{definition}

$J_{\C{E}_{\R{r}}} \ge 1$ corresponds to a fault condition we wish to avoid.
We now state the main problem under study in this paper. 

\begin{problem}[Minimax flow problem]\label{prb:minimax_flow}
For a flow network over an {\color{black}acyclic graph}, 
the \emph{minimax flow problem} is
\begin{equation}\label{eq:minimax}
  \begin{aligned}
     \min_{\B{m}_\R{s}}\ 
	 &\ J_{\C{E}_{\R{r}}}(\B{m}_\R{s}), \\
     \R{s.t.\ }\ 
	 &\begin{dcases}
			\B{B} \B{f} = \B{m},\\
			\textstyle \sum_{i \in \C{V}} m_i = 0,\\
    	\left| \B{f} \right| < \bar{\B{f}},\\
    	\B{m}_{\R{min}} \le \B{m}_\R{s} \le \B{m}_{\R{max}}.
	 \end{dcases}
  \end{aligned}
\end{equation}
\end{problem}

Following the steps in \cite{ahuja1986algorithms} and exploiting \eqref{eq:flows}, it is straightforward to verify that the minimax flow problem is a linear program
and can be solved using standard centralized iterative approaches. 
However, such an approach has two major drawbacks:
(i) it requires receiving data from all edges and transmitting data to all the suppliers, which can be impractical;
(ii) if $m_i, i \in \C{V}_\R{c}$ are time-varying, the optimization problem needs to be solved repeatedly and if the re-computation is not {\color{black}fast enough}, faults may occur from applying control inputs that are not up to date, as we will show in Section \ref{sec:simulations}.

As explained below, it might occur that the flow can be controlled only on a subset of the edges, say $\C{E}_{\R{cf}}$; therefore, in the rest of this paper, when considering Problem \ref{prb:minimax_flow} and the flow safety margin function $J_{\C{E}_{\R{r}}}$ in Definition \ref{def:flow_safety_margin}, we take ${\C{E}_{\R{r}}} = \C{E}_{\R{cf}}$, and omit the subscript of $J_{\C{E}_{\R{cf}}}$ (writing $J$), for the sake of brevity.
Next, we give a formal definition and characterization of the subset of edges with controllable flows $\C{E}_{\R{cf}}$.


\subsection{Edges with controllable flows}%
\label{sec:preliminary_concepts}

\begin{figure}[t]
\centering
\sidesubfloat[]{\includegraphics[max width=\columnwidth]{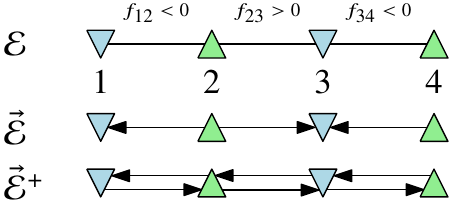}\label{fig:all_edges_set_a}}\\ \bigskip
\sidesubfloat[]{\includegraphics[max width=\columnwidth]{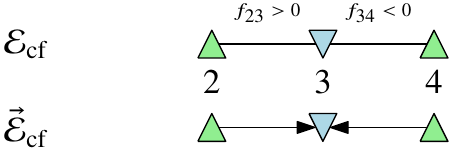}\label{fig:all_edges_set_b}}
\caption{{\color{black}(a), (b): The} various edges sets used in the paper, for a simple flow network. 
Upward green triangles represent suppliers, while downward blue triangles denote consumers.}
\label{fig:all_edges_set}
\end{figure}

Given an undirected graph $\C{G} = (\C{V}, \C{E})$ associated to a flow network, a \emph{set of directed edges} $\vec{\C{E}}$ is obtained by orienting the edges in $\C{E}$ according to the direction of the flows on them. Namely, for each $\{i,j\} \in \C{E}$, $\vec{\C{E}}$ contains either edge $(i,j)$ if $f_{ij} > 0$, or $(j,i)$ if $f_{ij} < 0$, or no edge if $f_{ij} = 0$.
We also define the \emph{extended set of directed edges} $\vec{\C{E}}^+$ as the set that, for each $\{i,j\} \in \C{E}$, contains both $(i,j)$ and $(j,i)$ (independently of the value of $f_{ij}$).
These sets are portrayed in Figure \ref{fig:all_edges_set_a}.

\floatsetup[figure]{style=plain,subcapbesideposition=top}

\begin{figure}[t]
\centering
\sidesubfloat[]{\includegraphics[max width=\columnwidth]{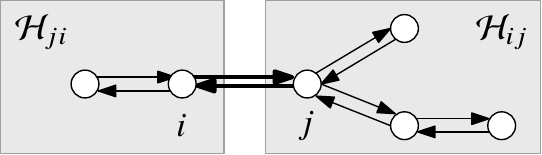}
\label{fig:half_clusters}}\\
\bigskip
\sidesubfloat[]{\includegraphics[max width=\columnwidth]{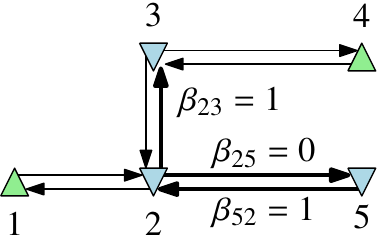}
\label{fig:microgrid_beta}}\\
\bigskip
\sidesubfloat[]{\includegraphics[max width=\columnwidth]{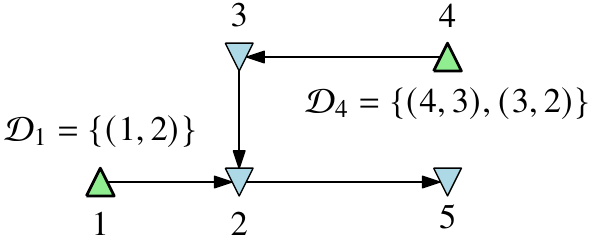}
\label{fig:downstreams}}
\caption{
(a): Half-clusters $\C{H}$ of edges $(j, i)$ (left) and $(i,j)$ (right) in an example graph $(\C{V}, \vec{\C{E}}^+)$.
(b): {\color{black}Supplier} indicator function $\beta$ for several edges, in an example graph $(\C{V}, \vec{\C{E}}^+)$; upward green triangles represent suppliers; downward blue triangles denote consumers.
(c): Some downstreams $\C{D}_i$ in an example graph $(\C{V}, \vec{\C{E}})$.}
\end{figure}

\begin{definition}[Half-cluster]
	For an acyclic undirected graph $\C{G} = (\C{V}, \C{E})$, the \emph{half-cluster} is a function $\C{H} : \vec{\C{E}}^+ \rightrightarrows \C{V}$.
	In particular, $\C{H}[(i, j)] = \C{H}_{ij}$ is the set of vertices in the connected component of $\C{G} \setminus \{i\}$ that contains $j$ (Figure \ref{fig:half_clusters}).
\end{definition}

\begin{definition}[Supplier indicator function]
	For an acyclic flow network, the \emph{supplier indicator function} $\beta : \vec{\C{E}}^+ \rightarrow \{0, 1\}$ is defined as
	\begin{equation}
		\beta[(i, j)] = \beta_{ij} \triangleq \begin{dcases}\label{eq:beta_ij}
			1, & \text{if} \ \C{V}_\R{s} \cap \C{H}_{ij} \ne \varnothing, \\
			0, & \text{otherwise}.
		\end{dcases}
	\end{equation}
\end{definition}
In simple terms, $\beta_{ij}$ is $1$ if a supplier can be be found in $\C{H}_{ij}$; moreover, notice that in general $\beta_{ij}$ is unrelated to $\beta_{ji}$.
A graphical example is given in Figure \ref{fig:microgrid_beta}.

\floatsetup[figure]{style=plain,subcapbesideposition=top}

As stated in the next Lemma, some flows $f_{ij}$ do not depend on the amount of commodity generated by supplier vertices, and thus we will not consider them in the optimization problem.
We define the \emph{set of edges with controllable flows} as
\begin{equation}\label{eq:E_r}
	\C{E}_{\R{cf}} \triangleq \{ \{i,j\} \in \C{E} \mid \beta_{ij} = 1 \wedge \beta_{ji} = 1\}.
\end{equation}

\begin{lemma}[Non-controllable flows]\label{lem:optimization_reduced_graph}
	{\color{black}In an acyclic flow network}, the flows $f_{ij}$ for $\{i, j\} \in \C{E} \setminus \C{E}_{\R{cf}}$ are independent of {\color{black}the supplied commodity} $m_k$, $\forall k \in \C{V}_\R{s}$.
\end{lemma}
\begin{proof}
	Consider an edge $\{i,j\} \in \C{E} \setminus \C{E}_{\R{cf}}$; by \eqref{eq:E_r}, it holds that $\beta_{ij} = 0 \ \vee \ \beta_{ji} = 0$.
	Without loss of generality, assume that $\beta_{ij} = 0$, which means that $\C{H}_{ij}$ contains no suppliers.
	Then, using \eqref{eq:flow_network} for all vertices in $\C{H}_{ij}$, we have that all edges reaching a vertex in $\C{H}_{ij}$ (including $\{i,j\}$) have their flows \emph{only} determined by ${\{m_q\}}_{q \in \C{H}_{ij}}$.
	As $\C{H}_{ij} \cap \C{V}_\R{s} = \varnothing$, we conclude that these flows do not depend on any $m_k$, for $k \in \C{V}_\R{s}$.
	\end{proof}


We define $\C{V}_{\R{cf}}$ as the set of vertices that are reached by at least an edge in $\C{E}_{\R{cf}}$, and the graph $\C{G}_{\R{cf}} = (\C{V}_{\R{cf}}, \C{E}_{\R{cf}})$.
It is immediate to verify that this graph (i) cuts out from $\C{G}$ the branches that contain only consumers, (ii) is connected, and (iii) all of its leaf vertices are suppliers.
Finally, we let $\vec{\C{E}}_{\R{cf}}$ be the set of directed edges obtained by orienting the edges in $\C{E}_{\R{cf}}$ according to the flows, similarly to what we did to obtain $\vec{\C{E}}$ from $\C{E}$.
Examples {\color{black}of} $\C{E}_{\R{cf}}$ and $\vec{\C{E}}_{\R{cf}}$ are depicted in Figure \ref{fig:all_edges_set_b}. 

\section{Consensus reformulation of the minimax flow problem}%
\label{sec:reformulation}
Next, we introduce the notions of maximum downstream flows and consumer clusters which will then be used to reformulate the minimax flow optimization problem (Problem \ref{prb:minimax_flow}) as a consensus problem. 

\begin{figure}[t]
	\centering
	\includegraphics[max width=\columnwidth]{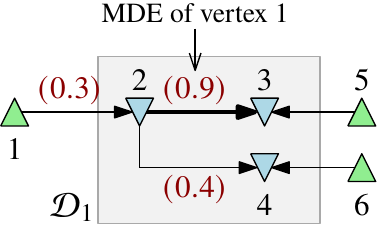}
	\caption{Representation of a maximum downstream edge (MDE). 
	Upward green triangles represent suppliers, while downward blue triangles denote consumers; in red and in parentheses we drew $f_{ij}/\bar{f}_{ij}$. 
	Edge $(2,3)$ is the MDE of vertex $1$. 
	Moreover, as vertex $1$ is a supplier, $(2,3)$ is a maximum downstream edge of a supplier vertex (MDES; i.e., $(2,3) \in \C{M}_{\R{s}}$), and, as vertex $3$ is a consumer, then also $(2,3) \in \C{M}_{\R{s} \rightarrow \R{c}}$.}
	\label{fig:maximum_downstream_edge}
\end{figure}

\begin{definition}[maximum downstream flows and edges]\label{def:downstream_definitions}
Consider a flow network associated to an acyclic graph $\C{G} = (\C{V}, \C{E})$.
Then,
\begin{enumerate}[(i)]
    \item\label{def:downstream} {\color{black}for} $i\in \C{V}$, the \emph{downstream} of vertex $i$, denoted by $\C{D}_i \subseteq \vec{\C{E}}_{\R{cf}}$, is the out-tree of vertex $i$ in $(\C{V}_{\R{cf}}, \vec{\C{E}}_{\R{cf}})$ ({\color{black}Figure} \ref{fig:downstreams});
    \item\label{def:Xi_definition} the \emph{maximum downstream flow} $\phi : \C{V} \rightarrow \BB{R}_{\ge 0}$ is given by
	\begin{equation}\label{eq:Xi_definition}
	\phi(i) = \phi_i \triangleq \max_{(j, k) \in \C{D}_i}\frac{f_{jk}}{\bar{f}_{jk}} \ge 0.
	\end{equation}
	\item\label{def:MDE_definition} {\color{black}for} $i\in \mathcal{V}$, the \emph{maximum downstream edge (MDE)} of vertex $i$ is $\arg \max_{(j, k) \in \C{D}_i} {f_{jk} / \bar{f}_{jk}} \in \vec{\C{E}}_{\R{cf}}$ ({\color{black}Figure} \ref{fig:maximum_downstream_edge}).
\end{enumerate}
\end{definition}

If $i \in \C{V}_\R{s}$, we abbreviate ``maximum downstream edge of a supplier vertex'' as \emph{MDES}. We denote by $\C{M}_\R{s} \subseteq \vec{\C{E}}_{\R{cf}}$ the set of all MDESs, and by $\C{M}_{\R{s}\rightarrow \R{c}} \subseteq \C{M}_\R{s}$ the set of MDESs that have consumers as terminal vertices (see again Figure \ref{fig:maximum_downstream_edge}).

We give next two instrumental results in Lemmas \ref{lem:equivalence_sets} and \ref{lem:critical_consumer_cluster}.

\begin{figure}[t]
	\centering
	\includegraphics[max width=\columnwidth]{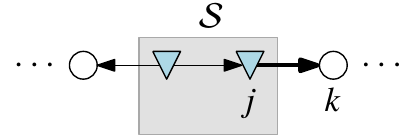}
	\caption{Graph topology described in the proof of Lemma \ref{lem:equivalence_sets}. Downward triangles represents consumers, while circles can either be suppliers or consumers.}
	\label{fig:lemma_equivalence_sets_edges}
\end{figure}

\begin{lemma}\label{lem:equivalence_sets}
	In an acyclic flow network, $\vec{\C{E}}_{\R{cf}} = \bigcup_{i \in \C{V}_\R{s}} \C{D}_i$.
\end{lemma}
\begin{proof}
	We obtain a proof by contradiction, showing that if the thesis did not hold, that would cause some consumer vertices not to receive as much commodity as they demand (which would contradict \eqref{eq:flow_network}). 
	In particular, contrary to the thesis, assume that there exists $(j, k) \in \vec{\C{E}}_{\R{cf}}$ such that 
	\begin{equation}\label{eq:proof_step_02}
		(j, k) \notin \bigcup_{i \in \C{V}_\R{s}} \C{D}_i.	
	\end{equation}
	Define $\C{S}$ as the set of vertices that have $(j, k)$ in their out-tree (see Figure \ref{fig:lemma_equivalence_sets_edges}).
	By Definition \ref{def:downstream_definitions}.\ref{def:downstream}, \eqref{eq:proof_step_02} implies that all nodes in $\C{S}$ are not suppliers (and thus are consumers), because the right-hand side in \eqref{eq:proof_step_02} is computed considering $i \in \C{V}_\R{s}$.
	Moreover, let $\C{E}_{\C{S}} \triangleq \{ (p,q) \in \vec{\C{E}}_{\R{cf}} \mid p \notin \C{S},  q \in \C{S}\}$ (i.e., edges ``on the boundary'' of $\C{S}$ that terminate in $\C{S}$).
	It is immediate to see that
	\begin{equation}\label{eq:proof_step_03}
		\C{E}_{\C{S}} = \varnothing;
	\end{equation}
	indeed, if there existed an edge $(p,q) \in \C{E}_{\C{S}}$, then $(j, k)$ would belong to the out-tree of $p$, which by definition of $\C{S}$ would imply $p \in \C{S}$, but this is impossible by definition of $\C{E}_{\C{S}}$.
	
	However, exploiting \eqref{eq:flow_network} for $i \in \C{V}_\R{c}$, we have that $\sum_{(j, k) \in \C{E}_\C{S}} f_{jk} = - \sum_{q \in \C{S}} m_q > 0$,%
	\footnote{\label{fn:consumers_not_supplied}Note that $\sum_{q \in \C{S}} m_q < 0$, rather than $\sum_{q \in \C{S}} m_q = 0$, because otherwise the vertices in $\C{S}$ would not be in $\C{V}_{\R{cf}}$.}
	which requires that $\C{E}_\C{S} \ne \varnothing$, but this is in contradiction with \eqref{eq:proof_step_03}.
	Therefore, an edge $(j, k)$ that satisfies \eqref{eq:proof_step_02} cannot exist, and the thesis is proved.
\end{proof}

\begin{figure}[t]
	\centering
	\sidesubfloat[]{\includegraphics[max width=\columnwidth]{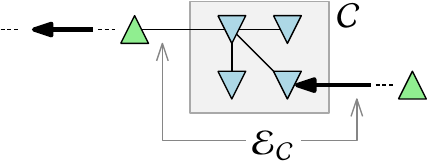}
	\label{fig:consumer_cluster}} \\ \bigskip
	\sidesubfloat[]{\includegraphics[max width=\columnwidth]{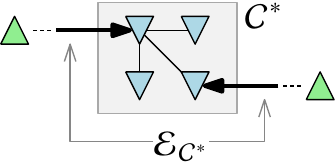}
	\label{fig:critical_consumer_cluster}} \\ \bigskip
	\sidesubfloat[]{\includegraphics[max width=\columnwidth]{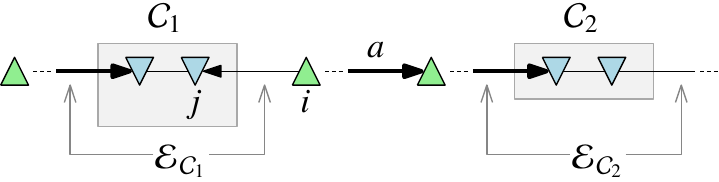}
	\label{fig:proof_supercritical_consumer_cluster}}
	\caption{(a): A consumer cluster $\C{C}$ (see Definition \ref{def:consumer_cluster}); upward green triangles represent suppliers, while downward blue triangles denote consumers; heavier arrows denote MDESs; dots represents connected components of vertices. 
	(b): A critical consumer cluster $\C{C}^*$ (see Lemma \ref{lem:critical_consumer_cluster}).
	(c): Situation described in the proof of Lemma \ref{lem:critical_consumer_cluster}.}
\end{figure}

\begin{definition}[Consumer cluster]\label{def:consumer_cluster}
	{\color{black}In an acyclic flow network}, a \emph{consumer cluster} $\C{C} \subset \C{V}_{\R{cf}}$ is a set of vertices having the following properties (see Figure \ref{fig:consumer_cluster}):
	\begin{enumerate}[(i)]
		\item \label{ite:all_are_consumers} all vertices in $\C{C}$ are consumers ($\C{C} \subseteq \C{V}_\R{c} \cap \C{V}_{\R{cf}}$), and $\C{C}$ is a connected component in $\C{G}_{\R{cf}} = (\C{V}_{\R{cf}}, \C{E}_{\R{cf}})$;
		\item \label{ite:no_MDES_inside} there are no MDESs between the vertices in $\C{C}$, i.e., $\C{M}_\R{s} \cap (\C{C} \times \C{C}) = \varnothing$;
		\item \label{ite:C_maximal_consumers} 
		any edge $(i,j)$ or $(j,i)$, where $i$ is a consumer not belonging to $\C{C}$ and $j$ is a vertex in $\C{C}$, must be an MDES; 
		\item \label{ite:exists_MDES} there exists at least an MDES that terminates in $\C{C}$, i.e., $\exists (i,j) \in \C{M}_{\R{s} \rightarrow \R{c}} : j \in \C{C}$.
	\end{enumerate}	
\end{definition}
Given a consumer cluster $\C{C}$, we denote by $\C{E}_\C{C} \subseteq \vec{\C{E}}_{\R{cf}}$ the set of directed edges that are on the boundary of $\C{C}$, i.e., $\C{E}_\C{C} \triangleq \{ (i,j) \in \vec{\C{E}}_{\R{cf}} \mid  (i \in \C{C}, j \not\in \C{C}) \vee (i \not\in \C{C}, j \in \C{C}) \}$.
Moreover, we denote by $\hat{\C{C}}$ the set of all consumer clusters and note the following facts.
Firstly, $\hat{\C{C}}$ is finite because the number of vertices in $\C{V}_{\R{cf}}$ is finite.
Secondly, any two different consumer clusters $\C{C}_1, \C{C}_2 \in \hat{\C{C}}$ must be disjoint, because of properties \ref{ite:no_MDES_inside}-\ref{ite:C_maximal_consumers} in Definition \ref{def:consumer_cluster}.
Thirdly, by Definition \ref{def:consumer_cluster}, any edge in $\C{M}_{\R{s}\rightarrow \R{c}}$ terminates in a consumer cluster.

\begin{definition}[Critical consumer cluster]\label{def:critical_consumer_cluster}
    {\color{black}In an acyclic flow network,} A \emph{critical consumer cluster} $\C{C}^*$ is a consumer cluster such that all $(i, j) \in \C{E}_{\C{C}^*}$ terminate in $\C{C}^*$ and are MDESs (see Figure \ref{fig:critical_consumer_cluster}), i.e.,
	\begin{equation}\label{eq:C_star}
		\forall (i,j) \in \C{E}_{\C{C}^*}, \quad (i,j) \in \C{M}_{\R{s} \rightarrow \R{c}} \wedge j \in \C{C}^*.
	\end{equation}
\end{definition}

\begin{lemma}[Existence of critical consumer cluster]\label{lem:critical_consumer_cluster}
	{\color{black}In an acyclic flow network,} if $\phi_i > 0$ for all $i \in \C{V}_\R{s}$, then there exists a critical consumer cluster.
\end{lemma}
\begin{proof}
	First, note that the hypothesis $\phi_i > 0, \forall i \in \C{V}_\R{s}$ implies that all suppliers have a MDE (that is a MDES; see Definition \ref{def:downstream_definitions}.\ref{def:MDE_definition}). 
	This, in conjunction with the facts that the network has an acyclic structure and that the number of vertices is finite, implies that 
	there exists at least a MDES terminating in a consumer, i.e., $\C{M}_{\R{s}\rightarrow \R{c}} \neq \varnothing$, which yields $\hat{\C{C}} \neq \varnothing$.

	Next, we prove the thesis by contradiction.
	{\color{black}Negating} the existence of a critical consumer cluster, we have, from  \eqref{eq:C_star},
	\begin{equation}\label{eq:existence_C_star_false}
		\forall \C{C} \in \hat{\C{C}}, \exists (i, j) \in \C{E}_{\C{C}} : \quad
		(i, j) \notin \C{M}_{\R{s} \rightarrow \R{c}} \vee j \notin \C{C}.
	\end{equation}
	
 	Let us consider some $\C{C}_1 \in \hat{\C{C}}$ and assume without loss of generality that the edge $(i,j)$ referenced in \eqref{eq:existence_C_star_false} is such that $i \notin \C{C}_1$ and $j \in \C{C}_1$ (i.e., $(i,j)$ ends in $\C{C}_1$; see Figure \ref{fig:proof_supercritical_consumer_cluster}).
 	In this case, it remains to be proved that assuming $(i, j) \notin \C{M}_{\R{s} \rightarrow \R{c}}$ leads to a contradiction.
	Indeed, in this case either $i$ is a supplier or it is a consumer. 
	In this latter case, by Definition \ref{def:consumer_cluster} (see in particular point \ref{ite:C_maximal_consumers}), $i$ must belong to $\C{C}_1$, which is against the hypothesis.
	If $i$ is a supplier instead, then it must have some MDES, say $a \in \C{M}_\R{s}$, that cannot be $(i,j)$ or belong to $\C{C}_1$ by Definition \ref{def:consumer_cluster} (point \ref{ite:no_MDES_inside}). 
	Then, either $a$ ends in a consumer or in a supplier.
	If it ends in a consumer, then $a$ must end in some consumer cluster $\C{C}_2$ different from $\C{C}_1$, given the property that the graph is acyclic by hypothesis.
	On the other hand, if $a$ ends in a supplier, then that supplier must have its own MDES and the argument can be repeated until an MDES ending in a consumer is found; hence, this MDES ends in a consumer cluster, which is different from any other defined earlier on in the procedure (because the graph $\C{G}$ {\color{black}is} acyclic).
	As this argument can be repeated \emph{ad infinitum}, we {\color{black}get} a contradiction (because $\hat{\C{C}}$ must be finite) and the theorem remains proved.	
	
	A similar argument could be used to reach a contradiction if the edge $(i, j)$ is assumed to be such that $i \in \C{C}_1$ and $j \notin \C{C}_1$ (i.e., $(i,j)$ does not end in $\C{C}_1$).
	Therefore, we conclude that \eqref{eq:existence_C_star_false} does not hold, which corresponds to the thesis.
\end{proof}

We are now ready to present our main result.

\begin{theorem}[Consensus achieves optimization]\label{thm:minimax_Xi}
{\color{black}In an acyclic flow network,} if $\phi_i = \phi^*$ for all $i \in \C{V}_\R{s}$ and for some $\phi^* \in \BB{R}_{\ge 0}$, then the cost function $J$ (see Definition \ref{def:flow_safety_margin}) is minimized with respect to $\B{m}_\R{s}$.
\end{theorem}

\begin{proof}
	From \eqref{eq:J}, exploiting Lemma \ref{lem:equivalence_sets}, and using \eqref{eq:Xi_definition}, we have 
	\begin{equation}\label{eq:J_as_Xi}
		J = \max_{\{ i,j \} \in \C{E}_{\R{cf}}} \frac{ \left| f_{ij} \right|  }{\bar{f}_{ij}} = \max_{(i,j) \in \vec{\C{E}}_{\R{cf}}} \frac{f_{ij}}{\bar{f}_{ij}} = 
		\max_{(i, j) \in {\bigcup_{i \in \C{V}_\R{s}} \C{D}_i}} \frac{ f_{ij} }{\bar{f}_{ij}} = \max_{i \in \C{V}_\R{s}} \phi_i.
	\end{equation}
	From \eqref{eq:J_as_Xi}, it is obvious that, if $\phi^* = 0$, then $J = 0$, which clearly corresponds to the lowest possible value of $J$. 

	We consider next the case that $\phi^* > 0$.
	For the sake of brevity, let $x_{ij} \triangleq f_{ij} / \bar{f}_{ij}$. 
	From Lemma \ref{lem:critical_consumer_cluster}, there exists a critical consumer cluster $\C{C}^*$, and using \eqref{eq:J_as_Xi} and the fact that $\C{E}_{\C{C}^*} \subseteq \vec{\C{E}}_{\R{cf}}$ we have
	\begin{equation}\label{eq:J_tilde}
		J \ge \tilde{J} \triangleq \max_{(i, j) \in \C{E}_{\C{C}^*}} x_{ij}.
	\end{equation}
	Then, from \eqref{eq:flow_network}, it is straightforward to compute that
	\begin{equation*}
		\sum_{(i, j) \in \C{E}_{\C{C}^*}} f_{ij} = - \sum_{k \in {\C{C}^*}} m_k,
	\end{equation*}
	which, letting $m_{\C{C}^*} \triangleq - \sum_{k \in {\C{C}^*}} m_k > 0$, can be rewritten as
	$\sum_{(i, j) \in \C{E}_{\C{C}^*}} x_{ij} \bar{f}_{ij} = m_{\C{C}^*}$.
	Therefore, considering the problem
	\begin{equation*}
		\begin{aligned}
		\min_{x_{ij} \in \BB{R}_{\ge 0}, (i,j) \in \C{E}_{\C{C}^*}} &\tilde{J},\\
		\R{s.t.\ } &\sum_{(i, j) \in \C{E}_{\C{C}^*}} x_{ij} \bar{f}_{ij} = m_{\C{C}^*},
		\end{aligned}
	\end{equation*}
	and recalling \eqref{eq:J_tilde}, it is clear that the minimum value of $\tilde{J}$ is achieved when all $x_{ij}$s are equal.
	At this point, by hypothesis, $x_{ij} = \phi^*, \forall (i, j) \in \C{E}_{\C{C}^*}$, and thus $\tilde{J} = \phi^*$ is minimal.
	From \eqref{eq:J_as_Xi} and the hypothesis, it also holds that $J = \phi^*$; therefore, from \eqref{eq:J_tilde}, $J$ is also minimized.
\end{proof}

Note that Theorem \ref{thm:minimax_Xi} offers only a sufficient condition for the solution of Problem \ref{prb:minimax_flow}.

\section{Distributed estimation of maximum downstream flows}%
\label{sec:estimation}

In this section, we study how the maximum downstream flows $\phi_i$ can be estimated by each node using a recursive process that only requires local information.
Then, in Section \ref{sec:application_to_microgrids}, we embed such estimation process in a heuristic distributed control approach to achieve consensus of the maximum downstream flows, and hence solve Problem \ref{prb:minimax_flow} via Theorem \ref{thm:minimax_Xi}, for the case of electric microgrids.

Let us denote by $\C{V}_i^{\R{out}}$ the out-neighborhood of vertex $i$ in the graph $(\C{V}, \vec{\C{E}})$.

\begin{lemma}[Reformulation of maximum downstream flows]\label{lem:Xi_recursive}
{\color{black}In an acyclic flow network,} the maximum downstream flow $\phi_i$ {\color{black}(see} Definition \ref{def:downstream_definitions}.\ref{def:Xi_definition}) can be found by computing 
\begin{equation}\label{eq:Xi_recursive}
\phi_i = \max_{j \in \C{V}^{\R{out}}_i} \left\lbrace
\beta_{ij} \frac{f_{ij}}{\bar{f}_{ij}}, \ 
\phi_{j} \right\rbrace.
\end{equation}
\end{lemma}

\begin{proof}
	
	\begin{figure}[t]
		\centering
		\includegraphics[max width=\columnwidth]{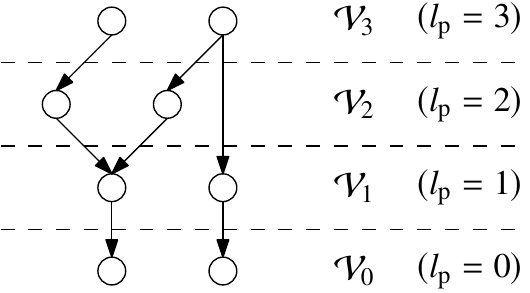}
		\caption{Grouping of vertices in accordance to their values of $l_\R{p}$, defined in the proof of Lemma \ref{lem:Xi_recursive}, for an example graph $(\C{V}, \vec{\C{E}})$.}
		\label{fig:graph_tree}
	\end{figure}
	
	For the sake of simplicity and without loss of generality, assume that $\bar{f}_{ij} = 1$ and $\beta_{ij} = 1$ for all $(i,j) \in \vec{\C{E}}^+$.
	In the directed acyclic graph $(\C{V}, \vec{\C{E}})$, let us denote by $l_\R{p}(i)$ the maximum length of all directed paths starting from vertex $i$; then $\C{V}_0, \C{V}_1, \C{V}_2, \dots$ are the sets of vertices that have $l_\R{p} = 0, l_\R{p} = 1, l_\R{p} = 2, \dots$, respectively (see Figure \ref{fig:graph_tree}).	
	We show the thesis, i.e., that \eqref{eq:Xi_recursive} is equivalent to \eqref{eq:Xi_definition}, for the subsets $\C{V}_0, \C{V}_1, \C{V}_2, \dots$ one at a time.
	\begin{itemize}
	
	\item $k \in \C{V}_0$. \
	As $\C{D}_k = \C{V}^\R{out}_k = \varnothing$, both \eqref{eq:Xi_definition} and \eqref{eq:Xi_recursive} yield
	\begin{equation}\label{eq:Xi_V_0}
		\phi_k = 0, \quad k \in \C{V}_0.
	\end{equation}
	
	\item $j \in \C{V}_1$. \
	We have $\C{D}_j = \{(j,k) \mid k \in \C{V}^\R{out}_j \}$.
	This, together with \eqref{eq:Xi_V_0}, means that both \eqref{eq:Xi_definition} and \eqref{eq:Xi_recursive} give
	\begin{equation}\label{eq:Xi_V_1}
	\phi_{j} = \max \left\{ f_{jk} \right\}_{k \in \C{V}^{\R{out}}_j},
	\quad j \in \C{V}_1.
	\end{equation}
	
	\item $i \in \C{V}_2$. \
	Now, $\C{D}_i = \{ (i,j) \mid j \in \C{V}^{\R{out}}_i\} \cup \{ (j,k) \mid j \in \C{V}^{\R{out}}_i, k \in \C{V}^{\R{out}}_j \}$,
	From \eqref{eq:Xi_definition}, we have
	\begin{equation}\label{eq:Xi_V_2}
	\phi_i = \R{max} \left\{ 
	{\left\{ f_{ij} \right\}}_{j \in \C{V}^{\R{out}}_i},
	{\left\{ f_{jk} \right\}}_{j \in \C{V}^{\R{out}}_i, k \in \C{V}^{\R{out}}_j}
	\right\},
	\quad i \in \C{V}_2.
	\end{equation}
	Then, using \eqref{eq:Xi_V_1}, \eqref{eq:Xi_V_2} can be rewritten as
	\begin{equation*}
	\phi_i = \R{max} \left\{ 
	{\left\{ f_{ij} \right\}}_{j \in \C{V}^{\R{out}}_i},
	{\left\{ \phi_{j} \right\}}_{j \in \C{V}^{\R{out}}_i}
	\right\},
	\quad i \in \C{V}_2,
	\end{equation*}
	which corresponds to \eqref{eq:Xi_recursive}.
	
	\item $h \in \{ \C{V}_3, \dots, \C{V}_{N-1} \}$. \
	The reasoning presented at the above point can be easily repeated to show that \eqref{eq:Xi_recursive} is equivalent to \eqref{eq:Xi_definition} for all remaining vertices.
	\end{itemize}
\end{proof}

In practice, the calculation in \eqref{eq:Xi_recursive} can be implemented through an arbitrarily fast dynamical estimation system, as stated in the next proposition.

\begin{proposition}[Distributed estimation of maximum downstream flows]\label{pro:Xi_hat_converges_to_Xi}
{\color{black}In an acyclic flow network,} we {\color{black}let} $\hat{\phi} : \C{V} \times \BB{R}_{\ge 0} \rightarrow \BB{R}$---denoting $\hat{\phi}(i, t)$ by $\hat{\phi}_i(t)$---be the solution to
	\begin{equation}\label{eq:Xi_estimation}
		\dot{\hat{\phi}}_i(t) = - k_\phi \left( 
		\hat{\phi}_i(t) - \max_{j \in \C{V}^{\R{out}}_i} \left\lbrace
		\beta_{ij} \frac{ f_{ij}}{\bar{f}_{ij}}, \
		\hat{\phi}_j(t) 
		\right\rbrace
		\right),
		\ \hat{\phi}_i(0) = 0,
	\end{equation}
$\forall i \in \C{V}$.
Assume {\color{black}the} $f_{ij}$s are constant, or $k_\phi \in \BB{R}_{>0}$ is large enough so that the $f_{ij}$s can be considered constant with respect to the dynamics of the $\hat{\phi}_i$s.
Then, $\hat{\phi}_i$ converges to $\phi_i$, $\forall i \in \C{V}$.
\end{proposition}
\begin{proof}
	As in the Proof of Lemma \ref{lem:Xi_recursive}, for {\color{black}simplicity} and without loss of generality, assume that $\bar{f}_{ij} = 1$ and $\beta_{ij} = 1$ for all $(i,j) \in \vec{\C{E}}^+$; moreover, consider again the sets $\C{V}_0, \C{V}_1, \C{V}_2, \dots$ defined in that Proof and depicted in Figure \ref{fig:graph_tree}.
	\begin{itemize}
		
		\item $k \in \C{V}_0$. \
		From \eqref{eq:Xi_estimation}, we have
		\begin{equation*}
			\dot{\hat{\phi}}_k(t) = - k_\phi \hat{\phi}_k(t),
			\quad \hat{\phi}_k(0) = 0,
			\quad k \in \C{V}_0.
		\end{equation*}
		Thus, for $k \in \C{V}_0$, $\forall t$, $\hat{\phi}_k(t) = 0 = \phi_k$ (see \eqref{eq:Xi_recursive}).

		\item $j \in \C{V}_1$. \
		From \eqref{eq:Xi_estimation} and what we stated at the previous point, we have
		\begin{equation}
			\dot{\hat{\phi}}_j(t) = - k_\phi \left( 
			\hat{\phi}_j(t) - \max_{k \in \C{V}^{\R{out}}_j} \left\{
			f_{jk}, 0 \right\} \right),
			\quad j \in \C{V}_1.
		\end{equation}
		Recall that all $f_{jk}$ can be considered constant by hypothesis.
		Therefore, $\forall j \in \C{V}_1$, $\hat{\phi}_j$ converges exponentially fast to $\phi_j$, as given in \eqref{eq:Xi_recursive}.

		\item $i \in \C{V}_2$. \
		From \eqref{eq:Xi_estimation}, we get
		\begin{equation}
			\dot{\hat{\phi}}_i(t) = - k_\phi \left( 
			\hat{\phi}_i(t) - \max_{j \in \C{V}^{\R{out}}_i} \left\{
			f_{ij}, \hat{\phi}_j(t) \right\} \right),
			\quad i \in \C{V}_2.
		\end{equation}
		After a short time, all $\hat{\phi}_j$, $j \in \C{V}_1$, can be considered at steady state. 
		Thus, clearly $\hat{\phi}_i$ converges to $\phi_i$ (as given in \eqref{eq:Xi_recursive}), for all $i \in \C{V}_2$.

		\item $h \in \{ \C{V}_2, \dots, \C{V}_{N-1} \}$. \						
		The above steps can be repeated to show the thesis for the remaining nodes.
	\end{itemize}
\end{proof}

To compute the generator indicator function $\beta$ appearing in \eqref{eq:Xi_estimation} (and defined in \eqref{eq:beta_ij}), we use the following algorithm, which ideally converges arbitrarily fast.
For each $(i,j) \in \vec{\C{E}}^+$, we define $\hat{\beta}_{ij}$, which is initialised to 1 if $j \in \C{V}_\R{s}$, or 0 otherwise.
Then, it is straightforward to verify that any $\hat{\beta}_{ij}$ converges exactly to $\beta_{ij}$ in at most $N-2$ steps, repeating the following Boolean assignments:
\begin{equation*}
\hat{\beta}_{ij} \leftarrow \hat{\beta}_{ij} \vee \left( \bigvee_{k \in \C{V} \mid k \ne i, (j, k) \in \vec{\C{E}}^+} \hat{\beta}_{jk} \right),
\quad \forall (i, j) \in \vec{\C{E}}^+.
\end{equation*} 

Next, we will show {\color{black}through} a representative application to microgrids that the distributed approach to estimate the maximum downstream flows can be used together with Theorem \ref{thm:minimax_Xi} to synthesize a heuristic control strategy able to solve the minimax flow optimization problem in a distributed manner.

\section{Application to microgrids}%
\label{sec:application_to_microgrids}
We consider an AC microgrid \cite{parhizi2015state} whose communication topology is described by an undirected, connected, acyclic, and weighted graph $\C{G} = (\C{V}, \C{E})$, with $N \triangleq |\C{V}|$ and $N_\C{E} \triangleq |\C{E}|$.
We let $\C{V}_\R{s} \triangleq (1, \dots, N_\R{s})$, where $N_\R{s} < N$, denote the set of power generators (suppliers), whereas $\C{V}_\R{c} \triangleq (N_\R{s} + 1, \dots, N)$ denotes loads (consumers).
We let $\C{Q}$ and $\B{B}$ be defined as in Section \ref{sec:review}.
Assuming (i) the generators are distributed energy resources with voltage source converters as power electronic interfaces, (ii) resistive loads, (iii) lossless lines, (iv) quasi-synchronization, and (v) constant voltages, the frequency dynamics can be described as \cite{simpson-porco2013synchronization,iyer2010generalized}:
\begin{subnumcases}{\label{eq:microgrid}}
	D_i \dot{\delta}_i(t) = 
	P_i - \sum\nolimits_{j=1}^{N} A_{ij} \sin (\delta_i(t) - \delta_j(t)), & $i \in \C{V}_\R{s}$,\label{eq:generators_dynamics}\\ 
	0 = P_i - \sum\nolimits_{j=1}^{N} A_{ij} \sin (\delta_i(t) - \delta_j(t)), & $i \in \C{V}_\R{c}$,\label{eq:loads_dynamics} 
\end{subnumcases}
where $\delta_i(t)$ is the voltage phase angle at node $i$ at time $t$;
$P_i$ is the power supplied or consumed at node $i$, with $P_i > 0$ if $i \in \C{V}_\R{s}$ and $P_i \le 0$ if $i \in \C{V}_\R{c}$;
$A_{ij} = E_{i} E_j \left\lvert Y_{ij} \right\rvert$, where $E_i$ is the voltage magnitude at node $i$ and $Y_{ij}$ is the admittance on the line between nodes $i$ and $j$ ($Y_{ij} = Y_{ji}$);
$D_i > 0$ is the droop coefficient of generator $i$;
$\xi_{ij}(t) = A_{ij} \sin(\delta_i(t) - \delta_j(t))$ is the power flow from $i$ to $j$ at time $t$.
Each edge $\{i,j\}$ can only bear a power flow equal (in absolute value) to $\bar{f}_{ij} \in \BB{R}_{>0}$ before breaking down or being disconnected.

For compactness, we also define
$\B{P} \triangleq [P_1 \ \cdots \ P_N]\T$,
$\B{P}_\R{s} \triangleq [P_1 \ \cdots \ P_{N_\R{s}}]\T$, 
$\B{D} \triangleq [ D_1 \ \cdots \ D_{N_\R{s}} \ 0 \ \cdots \ 0 ]\T \in \BB{R}^N$,
$\B{\xi}(t) \triangleq [ \xi_{i j}(t) ]\T_{(i,j) \in \C{Q}} \in \BB{R}^{N_\C{E}}$,
$\bar{\B{f}} \triangleq [ \bar{f}_{ij} ]\T_{(i,j) \in \C{Q}} \in \BB{R}^{N_\C{E}}$.


\subsection{Optimization problem}
\label{sec:microgrid_optimization_problem}
The asymptotic behaviour of \eqref{eq:microgrid} was characterised in \cite{simpson-porco2013synchronization} through the following theorem.

\begin{theorem}[Steady-state solution \cite{simpson-porco2013synchronization}]\label{thm:synchronous_solution}
	Let $\B{f} \in \BB{R}^{N_\C{E}}$ be defined implicitly by
 	\begin{equation}\label{eq:synchronous_solution}
 		\B{B} \B{f} = \B{P} - \omega \B{D},
 	\end{equation}
	where $\omega \triangleq (\sum_{i \in \C{V}} P_i) / (\sum_{i \in \C{V}_{\R{s}}} D_i)$.
	The following statements are equivalent:
	\begin{enumerate}[(i)]
		\item A unique locally stable phase-locked solution $\delta_1(t), \dots, \delta_N(t)$ of \eqref{eq:microgrid} exists such that $\lim_{t \rightarrow +\infty} \B{\xi}(t) = \B{f}$ and $\lim_{t \rightarrow +\infty} \dot{\delta}_i(t) = \omega$ for all $i \in \C{V}$;
		\item \label{ite:feasibility_synchronization} $\left| f_{ij} \right| / A_{ij} < 1$ for all $\{i,j\} \in \C{E}$.
	\end{enumerate}
\end{theorem}
We assume that in \eqref{eq:microgrid} the terms $A_{ij}$ are large enough that \ref{ite:feasibility_synchronization} in Theorem \ref{thm:synchronous_solution} holds.
Moreover, we highlight that \eqref{eq:synchronous_solution} is a flow network such as \eqref{eq:flow_network}, where $\B{m} = \B{P} - \omega \B{D}$, noting that $\sum_{i \in \C{V}} m_i = \sum_{i \in \C{V}} P_i - \omega \sum_{i \in \C{V}_\R{s}} D_i = 0$.
Therefore, to minimize the likelihood of line faults, we aim to regulate the power values $\B{P}_\R{s}$ in a distributed fashion so as to solve
\begin{equation}\label{eq:minimax_microgrid}
  \begin{aligned}
     \min_{\B{P}_\R{s}}\ 
	 &\ \max_{\{i,j\} \in \C{E}_{\R{cf}}} \frac{ \left| f_{ij} \right|  }{\bar{f}_{ij}}, \\
     \R{s.t.\ }\ 
	 &\begin{dcases}
		\B{B} \B{f} = \B{P} - \omega \B{D},\\
    	\left| \B{f} \right| < \bar{\B{f}},\\
    	\B{P}_{\R{min}} \le \B{P}_\R{s} \le \B{P}_{\R{max}},
	 \end{dcases}
  \end{aligned}
\end{equation}
which is a particularization of Problem \ref{prb:minimax_flow}, and where $\C{E}_{\R{cf}}$ is defined as in \eqref{eq:E_r}, and $\B{P}_{\R{min}}, \B{P}_{\R{max}} \in \BB{R}_{>0}^{N_\R{s}}$.

{\color{black}We remark that the problem in \eqref{eq:minimax_microgrid} does not aim at minimizing the economic cost of operation.
Therefore, if a network operator wishes to keep costs low, they might also alternate between cost-first strategies and prevention-first strategies, depending on the criticality of the current operating conditions, e.g., when the network is becoming particularly congested, or when some of the suppliers are shut down.}

\subsection{Heuristic distributed control approach}%
\label{sec:control_law}

Recall that in a flow network, according to Theorem \ref{thm:minimax_Xi}, Problem \ref{prb:minimax_flow} is solved if the maximum downstream flows $\phi_i$, $\forall i \in \C{V}_\R{s}$, achieve consensus.
We observed heuristically that this happens if (i) the suppliers' commodity ($m_i$) is taken as a function of time and varied continuously with the law
\begin{equation}\label{eq:generic_control_law}
    \dot{m}_i(t) = - k (\hat{\phi}_i(t) - \hat{\phi}_{\R{avg}}(t)), \quad \forall i \in \C{V}_\R{s},
\end{equation}
where $k \in \BB{R}_{>0}$ and $	\hat{\phi}_{\R{avg}}(t) \triangleq \R{mean} \left\{ \hat{\phi}_i(t) \right\}_{i \in \C{V}_\R{s}}$, and (ii) it holds that $\B{m}_\R{min} < \B{m}_\R{s}(t) < \B{m}_\R{max}$ at all time (see §~\ref{sec:review}).

On the basis of this observation, we let $P_i$, $i \in \C{V}_\R{s}$ (see \eqref{eq:microgrid}) be functions of time, and define the Boolean quantities
\begin{equation*}
	\gamma_{i}(t) \triangleq \begin{dcases}
		1, & \text{if } \ P_{\R{min},i} < P_i(t) < P_{\R{max},i},\\
		0, & \text{otherwise},
	\end{dcases}
	\qquad i \in \C{V}_\R{s};
\end{equation*}
we say that generator $i$ has \emph{saturated} if $\gamma_i = 0$.
We also define%
\footnote{{\color{black}The estimates $\hat{\phi}_i(t)$ are computed using the current values of the flows, i.e., replacing $f_{ij}$ with $\xi_{ij}(t)$ in \eqref{eq:Xi_estimation}.}
Moreover, in practice, $\hat{\phi}_\R{avg}$, $\hat{\phi}_{\R{avg},i}^{\text{n-sat}}$, and $\hat{\phi}_{\R{max},i}^{\text{sat}}$ can be estimated locally at the nodes through arbitrarily fast consensus protocols and simple information propagation schemes; e.g., see \cite{bullo2020lectures}.}
\begin{align*}
	\hat{\phi}_{\R{avg},i}^{\text{n-sat}}(t) &\triangleq \R{mean} \left\{ \{ \hat{\phi}_i(t) \} \cup \left\{ \hat{\phi}_j(t) \right\}_{j \in \C{V}_\R{s} \mid j \ne i, \gamma_j = 1 \ } \right\}, 
	&i \in \C{V}_\R{s},\\
	\hat{\phi}_{\R{max},i}^{\text{sat}}(t) &\triangleq \R{max} \left\{\hat{\phi}_j(t) \right\}_{j \in \C{V}_\R{s} \mid j \ne i, \gamma_j = 0}, 
	&i \in \C{V}_\R{s};
\end{align*}
in practice, $\hat{\phi}_{\R{avg},i}^{\text{n-sat}}$ is an average computed over non-saturated generators, always including $i$, whereas $\hat{\phi}_{\R{max},i}^{\text{sat}}$ is a maximum computed over saturated generators, always excluding $i$.
Omitting time dependence for the sake of brevity, we propose to select $P_i$, $\forall i \in \C{V}_\R{s}$, according to the law
\begin{subnumcases}{\dot{P}_i = \label{eq:control_law}}
	- k_P (\hat{\phi}_i - \hat{\phi}_{\R{avg}}), & $\text{if } \gamma_k = 1, \forall k \in \C{V}_\R{s}$, \label{eq:control_law_1} 
	\\
	\tilde{P}_i, & $\text{if } ( \exists k \in \C{V}_\R{s} : \gamma_k = 0 ) {} \wedge {}$ \notag \\
	& $\phantom{\text{if }} (\gamma_i = 1  \vee \zeta_i = 1),$ \label{eq:control_law_2} 
	\\
	0, & $\text{otherwise}$, \label{eq:control_law_3}
\end{subnumcases}
where $k_P \in \BB{R}_{>0}$, and, for $i \in \C{V}_\R{s}$,
\begin{align*}
\tilde{P}_i &\triangleq - k_P \left( \hat{\phi}_i - \hat{\phi}_{\R{avg},i}^{\text{n-sat}} \right) 
- k_P^\gamma \left( \hat{\phi}_{\R{avg},i}^{\text{n-sat}} - \hat{\phi}_{\R{max},i}^{\text{sat}} \right),\\
\zeta_i &\triangleq \begin{dcases}
	\begin{aligned}
		1, \vphantom{\tilde{P}_i} \\ \vphantom{\tilde{P}_i}
	\end{aligned} 
	& 
	\begin{aligned}
		\text{if } 
		&( P_i \le P_i^\R{min} \ \wedge \ \tilde{P}_i > 0 ) \ \vee  \\
		&( P_i \ge P_i^\R{max} \ \wedge \ \tilde{P}_i < 0 ), 
	\end{aligned}\\
	0, & \text{otherwise},
	\end{dcases}
\end{align*}
with $k_P^\gamma \in \BB{R}_{>0}$.
Note that $\zeta_i = 1$ if $i$ has saturated, but applying control law \eqref{eq:control_law_2} would bring $P_i$ closer to its admissible region (i.e., $P_{\R{min},i} < P_i < P_{\R{max},i}$).

{\color{black}
In \eqref{eq:control_law}, the main purpose of \eqref{eq:control_law_2} and \eqref{eq:control_law_3} is to factor in the constraint on power generation.
Indeed,}
when no generators have saturated, \eqref{eq:control_law_1} is active, resembling \eqref{eq:generic_control_law}, causing $\phi_i, \forall i \in \C{V}_\R{s}$ to converge (which solves \eqref{eq:minimax_microgrid} by virtue of Theorem \ref{thm:minimax_Xi}).
Nonetheless, if at least one generator saturates, \eqref{eq:control_law_2} becomes active.
In \eqref{eq:control_law_2}, the term $\hat{\phi}_i - \hat{\phi}_{\R{avg},i}^{\text{n-sat}}$ achieves convergence of $\phi_i, \forall i \in \C{V}_\R{s} : \gamma_i  = 1$ (non-saturated generators), whereas the term $\hat{\phi}_{\R{avg},i}^{\text{n-sat}} - \hat{\phi}_{\R{max},i}^{\text{sat}}$ reduces the gap between the $\phi_i$s of non-saturated generators and the $\phi_i$s of saturated ones.
Both effects decrease $\max_{i \in \C{V}_\R{s}} \phi_i$ as much as possible, thus achieving the optimum value of $J$ (see \eqref{eq:J}).
{\color{black}
To take into account more constraints or objectives, it might be required to further modify the control law.
}

\subsection{Numerical simulations}%
\label{sec:simulations}

\subsubsection*{Setup}
\begin{figure}[t]
\centering
\includegraphics[max width=\columnwidth]{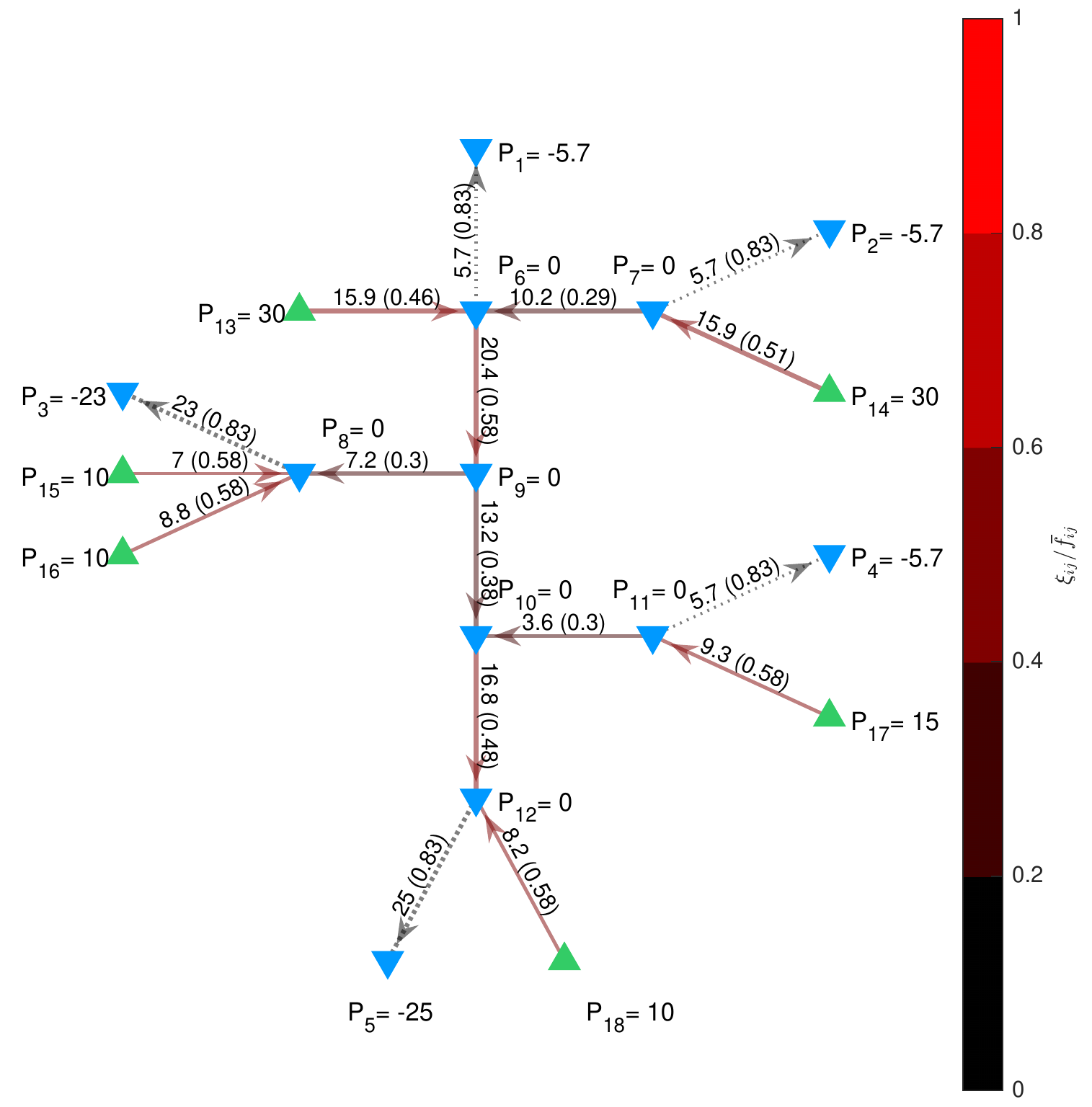}
\caption{Microgrid topology used in Section \ref{sec:simulations}, with active power values expressed in kW.
Upward green triangles are generators, while downward blue triangles are loads.
Dotted edges are those in $\C{E} \setminus \C{E}_{\R{cf}}$.
The values of the power flows $\xi_{ij}$ are the optimal ones with respect to \eqref{eq:minimax}, computed with the \textsc{Matlab} \texttt{minimax} function, and are reported on the edges. 
The fractions $\left\lvert \xi_{ij} \right\rvert / \bar{f}_{ij}$ are reported in brackets and the colors of the edges are a measure of proximity to failure.} 
\label{fig:scenario_1_graph}
\end{figure}

We tested our distributed estimation and control strategy {\color{black}\eqref{eq:Xi_estimation}-}\eqref{eq:control_law} on a benchmark problem and compared it to an offline centralized solution to \eqref{eq:minimax_microgrid}.
We used a slightly modified version of the standard CIGRE microgrid benchmark \cite{papathanassiou2005benchmark}, as depicted in Figure \ref{fig:scenario_1_graph}.
All computations were carried out in \textsc{Matlab} \cite{matlab2021version}; the centralized solution to \eqref{eq:minimax_microgrid} was found using the \texttt{fminimax} function; the parameters we used are $\B{P}_\R{min} = 0.8 \B{P}_\R{s}$, 
$\B{P}_\R{max} = 1.2 \B{P}_\R{s}$,
$k_\phi = 200$,
$k_P = 40$,
$k_P^\gamma = 40$.

\begin{figure}[t]
	\centering
	\includegraphics[max width=\columnwidth]{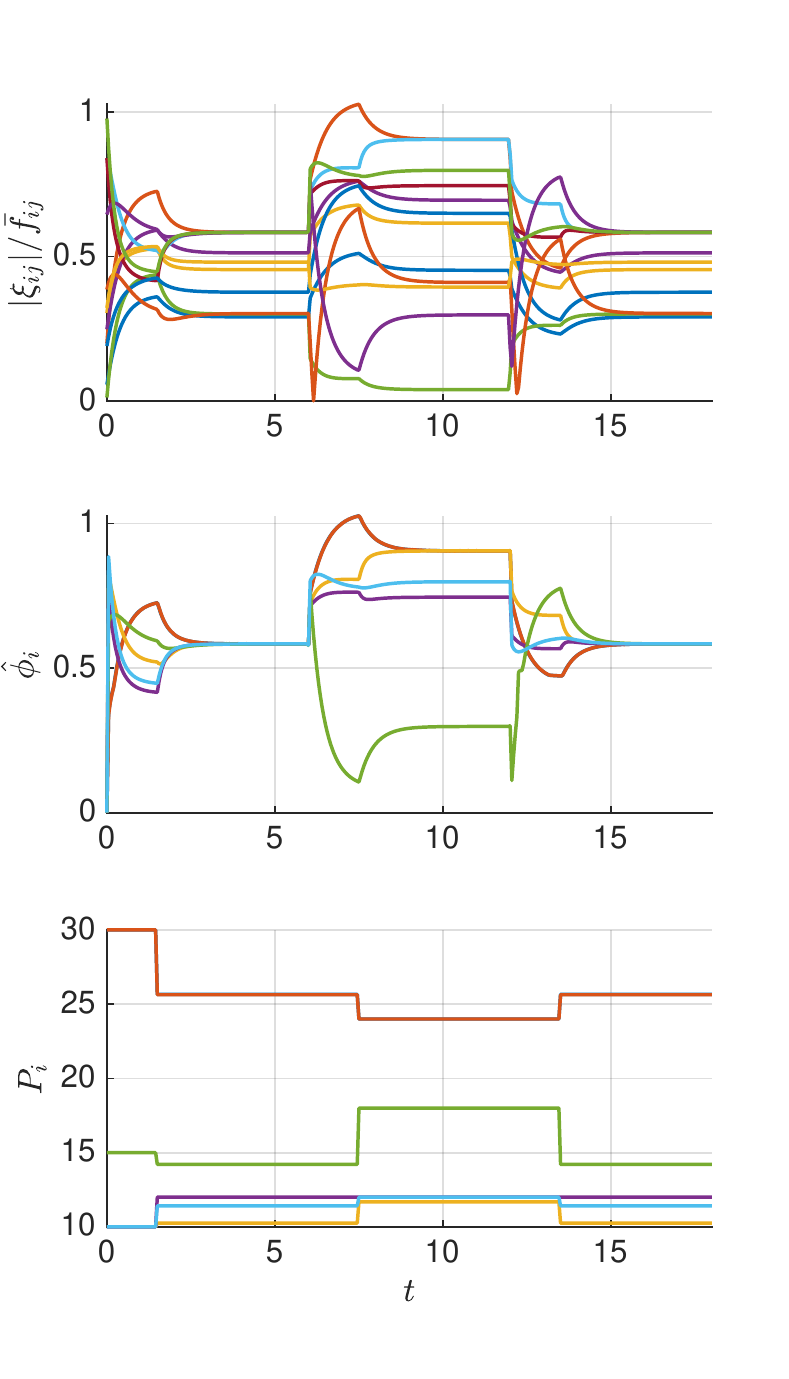}
	\caption{Results obtained when applying a centralized solution to \eqref{eq:minimax_microgrid}.
	In the top panel, different colors represent $|\xi_{ij}|/\bar{f}_{ij}$ for different edges, with $\{i,j\} \in \C{E}_{\R{cf}}$.
	In the middle and bottom panels, different colors represent $\hat{\phi}_i$ and $P_i$ for different supplier nodes, i.e., $i \in \C{V}_\R{s}$.}
	\label{fig:centralized_results}
\end{figure}

We simulated a scenario where the power values $P_i$ are initially assigned as in Figure \ref{fig:scenario_1_graph}; then, at time $t = 6$, $P_{9}$, $P_{10}$, $P_{11}$ become $-8$, $-4$, $-4$, respectively; at time $t = 12$, the original power values are restored.
These rapid fluctuations may represent the effect due to the plug-in and plug-out of multiple devices at once.
In Figure \ref{fig:centralized_results}, we report the results obtained by applying periodically an offline centralized solution to \eqref{eq:minimax_microgrid}. 
To account for the centralized and offline nature of this scheme, we consider a $1.5$ s delay in the application of the control values.
In Figure \ref{fig:distributed_results}, we show the results of applying our online distributed control strategy \eqref{eq:control_law}.
{\color{black}As a metric of performance, we consider $J^{\B{\xi}}(t) \triangleq \max_{\{i,j\} \in \C{E}_{\R{cf}}} \left\lvert \xi_{ij}(t) \right\rvert / \bar{f}_{ij}$; note that at steady state, when $\B{\xi} \rightarrow \B{f}$ (see Theorem \ref{thm:synchronous_solution}), we have $J^{\B{\xi}}(t) \rightarrow J$ (see § \ref{sec:minimax_problem}).}

\subsubsection*{Results}
For $0 \le t < 6$, at steady state, the optimal value $J = 0.584$ is obtained by both strategies.
In this time window, only \eqref{eq:control_law_1} is active, and convergence among all $\phi_i$, $i \in \C{V}_\R{s}$, is achieved, providing a practical demonstration of Theorem \ref{thm:minimax_Xi}.

For $6 \le t < 12$, the distributed control strategy achieves a maximum value {\color{black}(over time) of $J^{\B{\xi}}$ equal to} $0.915$, while the centralized scheme achieves $1.027$, which would trigger a fault ({\color{black}$J^{\B{\xi}} = 1$} is a fault condition).
{\color{black}This is an effect of the delay considered with this strategy to account for it being centralized and offline.}
At steady state, both strategies yield $J = 0.906$.
In this time window, several generators saturate; still, our distributed control strategy successfully achieves the optimal value of the cost function $J$, while preserving feasibility.

For $12 \le t \le 18$, both strategies yield the same optimal value of the cost function, that is $J = 0.587$.

\subsubsection*{Secondary controller}
\begin{figure}[t]
	\centering
	\includegraphics[max width=\columnwidth]{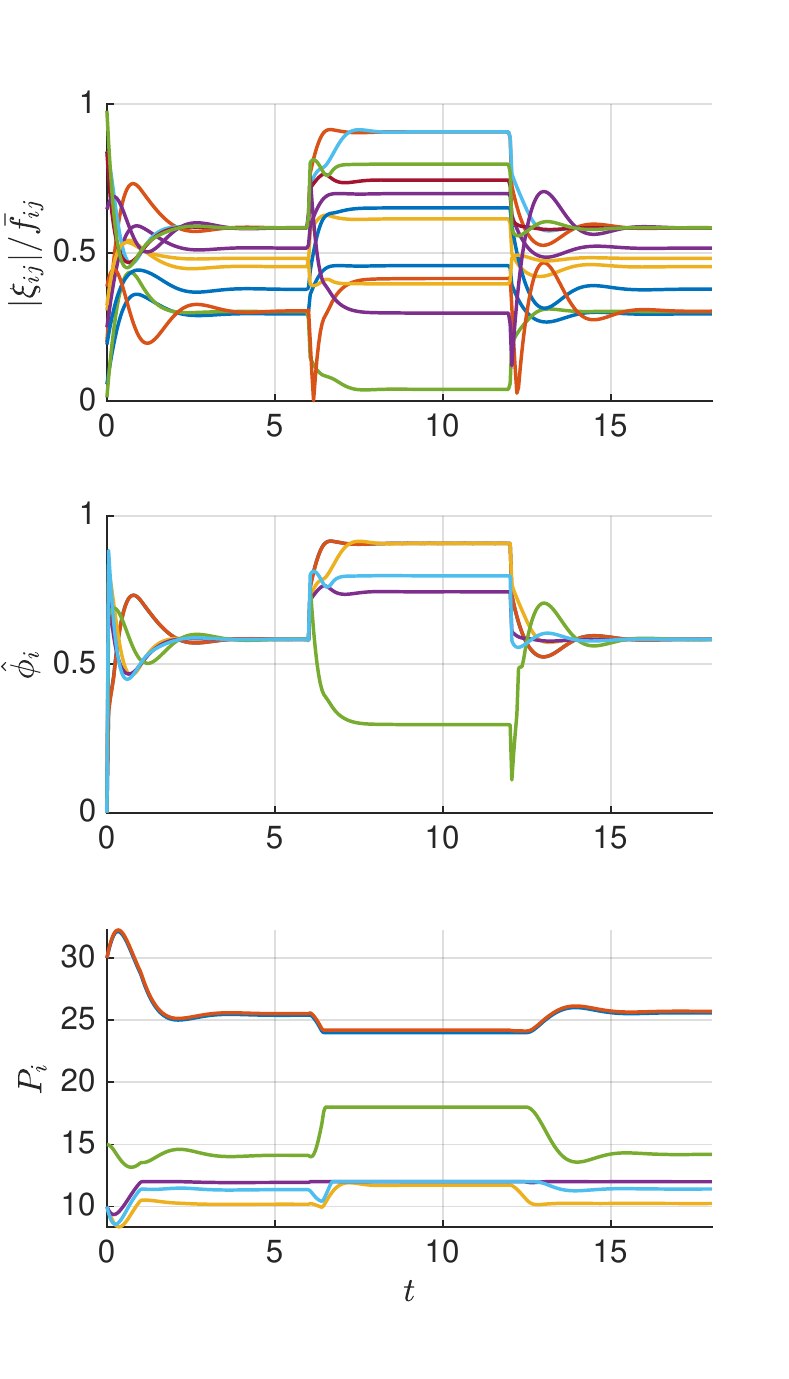}
	\caption{Results obtained when using the distributed online control strategy \eqref{eq:control_law}.}
	\label{fig:distributed_results}
\end{figure}

{\color{black}We also verified} that \eqref{eq:minimax_microgrid} can be solved by controlling $\B{D}$ (i.e., $D_i$s in \eqref{eq:generators_dynamics}), rather than $\B{P}_\R{s}$: this can be useful if one also wants to use a \emph{secondary controller} \cite[(16)]{simpson-porco2013synchronization} (to control $\B{P}_\R{s}$) with the aim to regulate the value of $\omega$ (defined in Theorem \ref{thm:synchronous_solution}).
In that case, \eqref{eq:control_law} is applied to $\dot{\B{D}}$, rather than to $\dot{\B{P}}_\R{s}$, {\color{black}and the} right-hand side of \eqref{eq:control_law} is multiplied by $-1$ (because $\B{D}$ appears with the minus sign in \eqref{eq:synchronous_solution}).
The results we obtain are qualitatively the same as those in Figure \ref{fig:distributed_results}, and thus we omit them here for brevity.


\section{Conclusion}\label{sec:conclusion}
We studied the minimax flow problem on acyclic networks showing that, by introducing the notion of maximum downstream flows, it can be reformulated as the problem of achieving their consensus.
We then proposed a distributed estimation strategy to evaluate maximum downstream flows.
We applied our results to the problem of preventing overcurrents in a droop-controlled AC microgrid via a distributed control strategy based on our approach. 
Our numerical experiments show that the distributed strategy is at least as effective, or even better, than the more traditional centralized solution strategy.

\subsubsection*{Extension to cyclic graphs}
Future research will address 
the extension of the approach to solve minimax flow problems on cyclic networks. This is particularly important in applications such as transmission grids where the network can have a meshed structure. 
{\color{black}%
In this paper, the assumption that the graph is acyclic (i) implies that the maximum downstream flow (MDF) of a supplier quantifies how much that node is contributing to network congestion, and (ii) is used to allow distributed computation of the MDFs.
Then, leveraging (i), the minimax flow problem is solved by balancing the MDFs.
The main challenge associated with 
extending the results presented here to cyclic graphs will be to design quantities analogous to the MDFs that satisfy these two properties.
}

\bibliographystyle{IEEEtran} 
\bibliography{references}

\end{document}